\documentstyle{article}
\def\be{\begin{equation}}
\def\ee{\end{equation}}
\def\bea{\begin{eqnarray}}
\def\eea{\end{eqnarray}}

\def\CP{{\cal PT}}
\def\p1{\psi_n(x_1)}
\def\p{\psi_n(x)}
\def\p2{\psi_n(x_2)}
\def\P{\Psi(t)}
\def\P0{\Psi(t_0)}
\def\P1{\Psi(x,t)}
\def\CT{{\cal T}}
\def\C{{\cal C}}

\def\Br{\Biggr}
\def\br{\biggr}
\def\Bl{\Biggl}
\def\bl{\biggl}
\def\l{\label}

\begin{document}

\hfill{}

\vskip 3
 \baselineskip
 \noindent

 {\Large\bf More on Exact $\CP$-Symmetric Quantum Mechanics}

\vskip \baselineskip

Khaled Saaidi{\footnote {E-mail: ksaaidi@uok.ac.ir,  \textrm{or},
KSaaidi@ipm.ir}

\vskip\baselineskip

{\small Department of Science,  University of Kurdistan, Pasdaran
Ave., Sanandaj, Iran } \\

 {\small Institute for Studies in Theoretical Physics and Mathematics,
 P.O.Box, 19395-5531, Tehran, Iran}

\vskip 2 \baselineskip

%{\bf PACS numbers}: 03.65.-w

%{\bf Keywords}:

\vskip 2\baselineskip

\begin{abstract}
In this article, we discussed certain properties of non-Hermitian
$\CP$-symmetry Hamiltonian, and it is shown that a consistent
physical theory of quantum mechanics can be built on a ${\cal C}
\CP$-symmetry Hamiltonian. In particular, we show that these
theories have unitary time evolution, and conservation probability.
Furthermore, transition from quantum mechanics to classical
mechanics is investigate and it is found that the Ehrenfest
theorem is satisfied.
\end{abstract}

\newpage
\section{Introduction}
It was shown that one can construct infinitely many new
Hamiltonian that  have $\CP$-symmetry  and it is rejected in past
because they are not Hermitian. In one the first explicit studies
of non-Hermitian Schr\"{o}dinger  Hamiltonians, have considered
the imaginary cubic oscillator problem in the context of
permutation theory \cite{cal} . It has recently been observed that
quantum mechanical  theories whose Hamiltonians are $\CP$-symmetry
have positive spectra even if the Hamiltonian is not Hermitian [2,
3, 4].  A class of such theories that has been studied extensively
is defined by the Hamiltonian,
 \be\l{1}
  H= p^2 - (ix)^N (N \geq 2).
 \ee

It is shown that the reality of the spectrum of (1) is due to
$\CP$-symmetry. Also, many quantum mechanical Hamiltonians have
been studied in great detail  by many authors[5-15]. Direct
numerical evidence for reality and positivity of the spectrum of
non-Hermitian  Hamiltonians, can be found  with great accuracy by
using conventional WKB approximation, and  performing a
Runge-Kutta integration of the associated Schr\"{o}dinger
equation[2, 4]. It is well known that the class of $\CP$-symmetry
Hamiltonian is larger than real Hamiltonian, because  all real
symmetric Hamiltonians have time reversal symmetric and one can
define the parity operator only up to unitary
equivalence\cite{b2}.

It is seen that the eigenfunctions  of a Hamiltonians H  which
have $\CP$-symmetry are simultaneously eigenfunctions of the
$\CP$ operator, if the $\CP$-symmetry of a Hamiltonian H is
not spontaneously broken\cite{b2}. In fact, even if the
Schr\"{o}dinger equation and the corresponding equation boundary
conditions are $\CP$-symmetric, the wave function that solve
the Schr\"{o}dinger equation boundary value problem may not be
symmetric under space-time reflection. When the solution exhibits
$\CP$-symmetry, the symmetry is unbroken, and if the solution
does not the $\CP$-symmetry, the $\CP$-symmetry is broken.

In [19, 20], it is shown  that $\CP$-symmetric Hamiltonians are
$\cal P$-pseudo Hermitian and, also it is found that the necessary
condition for having real spectrum is: \be \eta H \eta^{-1} =
H^{\dagger}, \ee where has been referred to a Hermitian linear
automorphism. One can, however, show that the real spectrum and
the orthognality of the states for some complex potentials can be
understood in terms of
 $\cal P$-pseudo Hermiticity of a Hamiltonian, provided the Hermitian
  linear automorphism, $\eta = e^{-\alpha p}$, which affects an imaginary
  shift of the coordinate ($x \rightarrow x + i \alpha $)\cite{Ahmed}. The author of
  \cite{mostafa1}, in some of his articles, attempt to demonstrate that
  $\CP$-symmetry can be understood from the theory of pseudo Hermitian operators
  and, in particular, he shown \cite{mostafa3}, that the exact $\CP$-symmetry
  is equivalent to Hermiticity. Also in \cite{23}, it is found that all
  $\CP$-symmetric quantum theories with real eigenvalues have the same
   completeness relation (5) which  for some especial example has been
   verified numerically in [10, 12].

The organization of this article is as follows. In sec.2, we offer
a discussion of ${\cal CPT}$-symmetry. In sec.3 and 4,  we discus
the conservation of probability,    the time evolution and
expectation value of an operator in non-Hermitian $\CP$-symmetry
quantum mechanics. In sec.5, we study the transition of quantum
mechanic to classical mechanic and it is shown that the Ehrenfest
theorem is satisfied.

\section{Preliminaries}
In this section we review a formal discussion of non-Hermitian
Hamiltonian which passes the $\CP$-symmetry, $[H, \CP] = 0$. In
quantum mechanic if a liner operator A commutes with H, then the
eigenstates  of H are also eigenstates  of  A. However, we know
that the operator  $\CP$ is anti-linear, but we emphasize that the
$\CP$-symmetry of H is unbroken, which means  that the
eigenfunction of H is simultaneously  an eigenfunction of $\CP$
[17, 18]. So that

 \bea\l{2}
  H\psi_n(x) &=& E_n\psi_n(x),\nonumber\\
  \CP\psi_n(x) &=& \lambda_n\psi_n(x).
  \eea
Using the facts that $(\CP)^2 = 1$ and ${\cal P}$ and $\CT$
commute with each other, also. We can conclude that $\psi_n(x) =
\lambda_n^* \lambda_n \psi_n(x)$, which means that $\lambda =
e^{i\alpha}$ for some  $\alpha \in \Re$. Thus it is seen that  one
can replace the  $\psi_n(x)$ by $e^{i\alpha/2}\psi_n(x)$, in which
its eigenvalue under the  operator $\CP$ is 1.
  \be\l{3}
  \CP\psi_n(x) = \psi_n(-x)^* = \psi_n(x).
  \ee
Whereas $[H, \CP]=0$, it is easily seen that the eigenvalue $E_n$
is real, $E_n=E_n^*$ [18]. The completeness relation of These
eigenfunctions, $\psi_n(x)$, in coordinate-space statement for
    all  $\CP$-symmetric quantum theories with real eigenvalues is \cite{23}.

 \be\l{4}
 \sum_n(-1)^n\psi_n(x_1)\psi_n(x_2) = \delta(x_1-x_2),
 \ee
for all $x,y \in \Re$. This completeness relation has been
verified numerically with great accuracy for some particular cases
of $\CP$-symmetric Hamiltonians in [10, 12]. Introducing the
${\cal PT}$ inner product for functions f(x) and g(x) as:

 \be\l{5}
 <f(x)|g(x)>_{\CP} := \int_c dx[\CP f(x)]g(x),
 \ee
where $c$ is an infinite contour in the complex-$x$ plane, which
the boundary conditions on the eigenfunctions are that
$\psi(x)\rightarrow 0$ exponentially rapidly as
$|x|\rightarrow\infty$ on the contour. It is clearly seen that the
eigenfunction are normalized to 1  as follows
 [17]
 \be\l{6}
 <\psi_n(x)|\psi_m(x)>_{\CP}=(-)^n\delta_{mn}, \ee
 where shows that the $\CP$-norm is not positive definite. It is found
that in any quantum mechanical  theory having an unbroken
$\CP$-Symmetry, there exists a symmetry of the Hamiltonian
connected with the fact that there are equal number of
positive-norm and negative-norm states. For this purpose, the
authors of\cite{br}, construct a linear operator, which denoted by
${\cal C}$, as:
 \be\l{7}
 {\cal C}(x_1, x_2) = \sum_n\psi_n(x_1)\psi_n(x_2), \ee
 where $\psi_n(x)$ is an eigenfunction of H. The operator
$\C$ has properties such as: i) $\C^2 =1$, ii) the inner product is
defined as:
  \be\l{8}
 <f(x)|g(x)>_{\cal CPT} := \int_c dx[{\cal CPT} f(x)]g(x),
 \ee
 whose associated norm is positive, because $\C$ contribute 1(-1)
when it acts on states with positive(negative) $\CP$ norm, and the
completeness relation (5), rewritten as:
  \be
  \sum_n [{\cal CPT}\psi_n(x)] \psi_n(y) = \delta(x - y),
  \ee
  which is valid for all $\CP$-symmetric Hamiltonians [22, 23].

\section{Conservation of Probability}

The time dependence Schr\"{o}dinger   equation of one dimensional
non-relativistic quantum theory is:

\be\l{9}
 i\hbar\frac{\partial \P1}{\partial t} = H\P1, \ee
 and for  the ${\cal CPT}$-symmetry Hamiltonian, H, we have

 \be\l{10}
            H(x)=H^*(-x).
            \ee
According to Born$^,$s postulate (one of the postulate of
Hermitian quantum mechanical theory), we define the probability of
finding the particle at time $t$ within the volume element $dx$
about the point $x$ for the case which the Hamiltonian is ${\cal
CPT}$-symmetry  as: \be\l{11}
 P(x, t) dx = [{\cal CPT}\P1]\P1 dx. \ee
Here we assume that the particle is described by a wave function
 $\P1$. By using of (7) and (10) it is seen that:

\be\l{12} \int_cP(x,t) dx = 1, \ee
where $c$ is the contour in complex-$x$ plane. It is clear that,
this probability is conserved, and at any arbitrary time the wave
function is normalized. In other hand, it is found that:

\bea\l{13} \frac{\partial}{\partial t}\int_cP(x,t)dx &=&
\frac{\partial}{\partial t}\int_c[{\cal CPT}\P1]\P1dx, \nonumber
\\
&=& \int_c {\Br (}\frac{\partial}{\partial t}[{\cal CPT}\P1]{\Bl
)}\P1dx + \int_c[{\cal CPT}\P1]\frac{\partial \P1}{\partial t}dx.
\eea Using(11) and its ${\cal CPT}$-Conjugate, we have

\be\l{14} -i\hbar\frac{\partial}{\partial t}{\Br [}{\cal CPT}
\P1{\Bl ]} = H^*(-x){\Br [}{\cal CPT }\P1{\Bl ]}. \ee Substitute
(16) and (11) in (15) and using this fact that $\psi(x)\rightarrow
0 $ exponentially rapidly as $|x|\longrightarrow\infty$ on the
contour, we have

\be\l{15}
\frac{\partial}{\partial t}\int_c P(x,t)dx+ \int_c
\frac{\partial j}{\partial x}dx = 0,
\ee
where \be\l{16} j(x, t) = \frac{\hbar}{2mi}{\Br \{}{\br [}{\cal
CPT} \P1{\bl ]}\frac{\partial \P1}{\partial x}
-{\Br (}\frac{\partial}{\partial x}{\br [}{\cal CPT} \P1{\bl ]}{\Bl )} \P1
 {\Bl\}}
 \ee
is the probability current density in ${\cal CPT}$-symmetric
quantum  theory, which for stationary states is independent of $x$
and $t$.
\section {The Time Evolution and Expectation Values}

Since the non-relativistic quantum  theories with any arbitrary
Hamiltonians is a first-order differential equation in time (11),
so one may introduce an evolution operator such as:

\be\l{17} \Psi(t) = U(t,t_0)\Psi(t_0), \ee where $U(t,t_0)=1$ and
$\Psi(t_0)$     is an initial wave function belonging to the
Hilbert space with the  ${\cal CPT}$ inner product and spanned by
the energy eigenfunctions.  Conservation of probability requires
that: \bea\l{18}<\Psi(t)|\Psi(t) >_{\cal CPT}
&=&  <\Psi(t_0)|\Psi(t_0) >_{\cal CPT}, \nonumber \\
&=&  <\Psi(t_0)|U^\sharp(t,t_0)U(t,t_0)|\Psi(t_0)>_{\cal CPT},
\eea
hence
\be\l{19}
 U^\sharp(t,t_0)U(t,t_0) = I,
 \ee
where $U^\sharp(t,t_0)$ is ${\cal CPT}$-conjugate of $U(t,t_0)$.
This shows that

\be\l{20} U^\sharp(t,t_0) = U^{-1}(t,t_0). \ee By instituting (19)
in (11), it is found that:

 \be\l{21}
 i\hbar\frac{\partial}{\partial t}U(t,t_0) = H U(t,t_0).
 \ee
Let us consider the particular case for which $\frac{\partial H
}{\partial t}=0$.
 A solution of (23), which satisfying the initial condition is given
by:

\be\l{22} U(t,t_0) = e^{-i/\hbar H (t-t_0)},\ee
where

\bea\l{23}
U^\sharp(t,t_0)&=&e^{i/\hbar H^*(-x) (t-t_0)},\nonumber \\
&=& e^{i/\hbar H (t-t_0)},\nonumber\\
&=&U^{-1}(t,t_0), \nonumber\\
&=& U(t_0, t). \eea
Thus a formal solution of the time-dependent Schr\"{o}dinger
equation for a time independent ${\cal PT}$-symmetry  Hamiltonian
is given by
\be\l{24}
 \Psi(t) = e^{-i/\hbar H(t-t_0)}\Psi(t_0).
 \ee

\section {Transition From Quantum Mechanics to Classical
Mechanics}

 According to the principle of correspondence, we aspect
that the motion of a wave packet should agree with that of the
corresponding classical particle whenever the distances and
momenta involved in describing the motion of the particle are so
large that uncertainty principle may be ignored.  Let us $\hat{A}$
be a linear operator which associated with a
  dynamical variable in this version of quantum theory. It acts in a complex
   vector space. For the case that $\hat{A}$
    is a proper physical  quantity, we want
   the $<\hat{A}>$  or the eigenvalues of it be real. We call this quantity as physical
   observable of theories.
   So the reality of $<\hat{A}>$ or eigenvalues of it follows that, $\hat{A}$ commute with
   nonlinear invertible operator $\CP$ [19, 20]. Therefore, it is necessary that
   $\hat{A}$ has been an un-broken $\CP$-symmetry. Also with pseudo
   operators points of view, it is necessary that $\hat{A}$ be a pseudo operator,
   which is related to a Hermitian operator $\cal A$ by a similarity
   transformation, as
   \be
   \hat{A}    = \eta^{-1} {\cal A}\eta,
   \ee
   where $\eta$ is an linear invertible operator in Hilbert space of the
    quantum system. Hence, whereof, the $\CP$-symmetric quantum theory
    is $\cal P$-pseudo hermitian quantum theories [19, 20] and these theories
    (exact $\CP$-symmetry) is equivalent to a Hermitian quantum theories, so we
     may introduce a proper observable  of $\CP$-symmetric quantum theory as similar
      transformation of the observable of a Hermitian quantum theory,
      in which,  one can defined that as pseudo observable. The expectation
value of this observable in the state $|\Psi>$, normalized to
unity, for a ${\CP}$-symmetry  quantum mechanical theory, is
define as follows:
 \be\l{25}
 <\hat{A}>_{\cal CPT} = <\Psi | \hat{A} | \Psi>_{\cal CPT},
 \ee
where $<|$ is ${\cal CPT}$-conjugate of $|>$. The rate of change
of this expectation value is therefore \bea\l{26}
\frac{d}{dt}<\hat{A}>_{\cal CPT}&=&\frac{d}{dt}<\Psi|\hat{A}|\Psi>_{\cal CPT}, \\
 &=&<{\partial \Psi \over \partial t}|\hat{A}|\Psi>_{\cal CPT} +<\Psi|{\partial
\hat{A} \over \partial t}|\Psi>_{\cal CPT} +
<\Psi|\hat{A}|{\partial \Psi \over \partial t}>_{\cal CPT},
 \eea
 then,
one can use of (11), (12) and (16)  arrive at:
 \be\l{27} {d \over
dt}<\hat{A}>_{\cal CPT}= (\frac{-1}{i\hbar})<H
\Psi|\hat{A}|\Psi>_{\cal CPT} +<\Psi|\frac{\partial
\hat{A}}{\partial t}|\Psi>_{\cal CPT} +
(\frac{1}{i\hbar})<\Psi|\hat{A}H|\Psi>_{\cal CPT}. \ee Since H is
a ${\CP}$-symmetry, H=H$^*$(-x), so  the first matrix element on
the right hand of (31) may be written as $<\Psi|HA|\Psi>$.
Therefore we have \be\l{29} {d \over dt}<\hat{A}>_{\cal CPT}=
(\frac{i}{\hbar})<[H ,\hat{A}]>_{\cal CPT}
 +<\frac{\partial \hat{A}}{\partial t}>_{\cal CPT}.
 \ee
 We assume that $\hat{A}=\hat{x}$,
       where is a position operator in $\CP$-symmetry quantum theory
      that acts in a complex vector space, which it may be a similarity
      transformation of $x_{\small\textrm{h}}$ as $\hat{x} = e^{-p\alpha} x_{\small\textrm{h}}
      e^{p\alpha} = x_{\small\textrm{h}} + i\alpha$ ($\alpha $=
      constant and $\small\textrm{h} :\equiv \textrm{Hermitian}$).
 So $\frac{\partial \hat{x}}{\partial t}=0$ and also
it is clearly seen that for any ${\CP}$-Symmetry Hamiltonian such
as (11), the commutative $[H, \hat{x}]$ is equal to: \be\l{30}
                [H,\hat{x}]=-i\hbar{\hat{p} \over m}.
\ee Then, from (29), \be\l{31} {d \over dt}<\hat{x}>_{\cal CPT} =
{<\hat{p}>_{\cal CPT} \over m}, \ee and also for the case
$\hat{A}=\hat{p}$, in which, $\hat{p} = e^{-p\alpha}
p_{\small\textrm{h}}
      e^{p\alpha} = p_{\small\textrm{h}} $,  one can arrive at:
 \be\l{32} {d \over
dt}<\hat{p}>_{\cal CPT} =- <{\partial \hat{V} \over \partial
\hat{x}}>_{\cal CPT} . \ee
 These two equations ((34) and (35)) show
that in non-Hermitian ${\CP}$-symmetry quantum mechanics, the
Ehrenfest theorem is satisfied exactly such as standard
formulation of quantum mechanics in term of Hermitian
Hamiltonians.

\section{Conclusion}.
During the past five years there have appeared dozens of
publications exploring the properties of the ${\CP}$-symmetric
Hamiltonians. It is found that a consistent physical theory of
quantum mechanics can be built on a non-Hermitian Hamiltonian,
which satisfies the lees restrictive and more physical condition
of space-time reflection symmetry. Considering some aspect of
these theories, it is shown that these theories are a consistence
physical quantum mechanic theory. In particular it is found that
the time-evolution of these theories is unitary and the
probability is conserved. Also by defining the expectation value
of observable with respect to ${\cal CPT}$-inner product, it is
shown that the rate of expectation value is similar to rate of
expectation value in standard Hermitian quantum mechanics. At
last, by considering the Ehrenfest theorem, it is seen that
transition from ${\CP}$-symmetric quantum mechanic to classical
mechanic is satisfied.

\end{document}